\documentclass[aps,pre,epsfig,epsf]{revtex}
\usepackage{multicol}
\begin{document}
\newcommand{\be}{\begin{equation}}
\newcommand{\ee}{\end{equation}}
\newcommand{\ba}{\begin{eqnarray}}
\newcommand{\ea}{\end{eqnarray}}

\title
{Analysis of  Self-Organized Criticality in the
Olami-Feder-Christensen model and in real earthquakes}

\author
{F. Caruso$^{1} $, A. Pluchino$^{2}$, V. Latora$^{2}$, S.
Vinciguerra$^{3}$, A. Rapisarda$^{2}$}

\address
{$^{1}$ NEST CNR-INFM \& Scuola Normale Superiore, Piazza dei
Cavalieri 7, I-56126 Pisa and\\Scuola Superiore di Catania,
Universit\`a di Catania, Via S. Nullo 5/i, I-95123 Catania, Italy\\
$^{2}$Dipartimento di Fisica e Astronomia, Universit\`a di
Catania, \\ and INFN sezione di Catania, Via S. Sofia 64, I-95123
Catania, Italy\\
$^{3}$HP-HT Experimental Laboratory of Volcanology and Geophysics,
Dept. of Seismology and Tectonophysics, \\
INGV, I-00143 Rome, Italy}

\maketitle

\begin{abstract}
We perform a new  analysis on the dissipative
Olami-Feder-Christensen model on a small world topology
considering avalanche size differences. We show that when criticality
appears the
Probability Density Functions (PDFs) for the avalanche size
differences at different times have fat tails with a q-Gaussian
shape.
This behaviour does not
depend on the time interval adopted and is found also
when considering energy differences between real earthquakes.
Such a result can be analytically understood if the sizes
(released energies) of the avalanches (earthquakes) have no
correlations.
   Our findings support the hypothesis that a
self-organized criticality mechanism with long-range interactions
is at the origin of seismic events and indicate  that it is not
possible to predict the magnitude of the next earthquake knowing
those of the previous ones.
\end{abstract}


\begin{multicols}{2}

In the last years there has been an intense debate on earthquake
predictability \cite{debates} and a  great effort in studying
earthquake triggering and interaction
\cite{marsan,casarot,cresce,parsons}. Along these lines the
possible application of the Self-Organized Criticality (SOC)
paradigm
\cite{BTW,bak_book,jen_book,bak,ma,com-corral,Olami,lise,sornette}
has been  discussed. Earthquakes trigger dynamic and static stress
changes. The first acts at short time and spatial scales,
involving the brittle upper crust, while the second involves
relaxation processes in the asthenosphere and acts at long time
and spatial scales
\cite{kagan,turcotte,palatella,abe,corral1,tosi,varotsos}. In this
letter, by means  of a new analysis, we show that it is possible
to reproduce statistical features of earthquakes catalogs
\cite{wc,nc} within a SOC scenario taking into account long-range
interactions. We consider the  dissipative Olami-Feder-Christensen
model \cite{Olami} on a \textit{small world} topology
\cite{Filippo_creta,watts} and we  show that the Probability
Density Functions (PDFs) for the avalanche size differences at
different times have fat tails with a q-Gaussian shape
\cite{tsallis,frisch,stella,beck} when finite-size scaling is
present. This behaviour does not depend on the time interval
adopted and is  found also when considering energy differences
between real earthquakes. It is possible to explain this result
analytically assuming the absence of correlations among the sizes
(released energies) of  the avalanches (earthquakes). This finding
does  not allow to predict the magnitude of the next earthquake
knowing those of the previous ones.

The Olami-Feder-Christensen (OFC) model\cite{Olami} is one of the
most interesting models displaying Self-Organized Criticality.
Despite of its simplicity, it exhibits a rich phenomenology
resembling real seismicity, like the presence of aftershocks  and
foreshocks \cite{sornette}. In its original version the OFC model
consists of a two-dimensional square lattice of $N=L^2$ sites,
each one connected to its $4$ nearest neighbours and carrying a
seismogenic force represented by a real variable $F_i$, which
initially takes a random value in the interval $(0,F_{th})$. In
order to mimic a uniform tectonic loading all the forces are
increased simultaneously and uniformly, until one of them reaches
the threshold value $F_{th}$ and becomes unstable $(F_i \geq
F_{th})$. The driving is then stopped and an ''earthquake'' (or
avalanche) starts:
\begin{equation}
 \label{av_dyn}
     F_i \geq F_{th}  \Rightarrow \left\{ \begin{array}{l}
      F_i \rightarrow 0 \\
      F_{nn} \rightarrow F_{nn} + \alpha F_i
\end{array} \right.
\end{equation}
where ''nn'' denotes the set of nearest-neighbor sites of $i$. The
number of topplings during an avalanche defines its size $S$,
while the dissipation level of the dynamics is controlled by the
parameter $\alpha$. The model is conservative if $\alpha=0.25$,
while it is dissipative for $\alpha<0.25$. In the present letter
we consider the dissipative version of the OFC model with
$\alpha=0.21$ \cite{footnote}, on a regular lattice with $L=64$
and open boundary conditions (i.e. we impose $F=0$ on the boundary
sites). But, in order to improve the model in a more realistic
way, we introduce a small fraction of long-range links in the
lattice so to obtain a small world topology \cite{watts}. Just  a
few long-range edges create short-cuts that connect sites which
otherwise would be much further apart. This kind of structure
allows the system to synchronize and to show both finite-size
scaling and universal exponents \cite{Filippo_creta}. The curves
obtained for different sizes of the system collapse into a single
one. Furthermore, a small world topology is expected to model more
accurately earthquakes spatial correlations, taking into account
long-range as well as short-range seismic effects
\cite{marsan,casarot,cresce,parsons}. In our version of the OFC
model the links of the lattice are rewired at random with a
probability $p$ as in the one-dimensional model of
Ref.\cite{watts}. In \cite{Filippo_creta} it was shown that the
transition to obtain small world features and criticality is
observed at  $p=0.02$.
%
\begin{figure} [ht]
\begin{center}
\includegraphics[width=8cm]{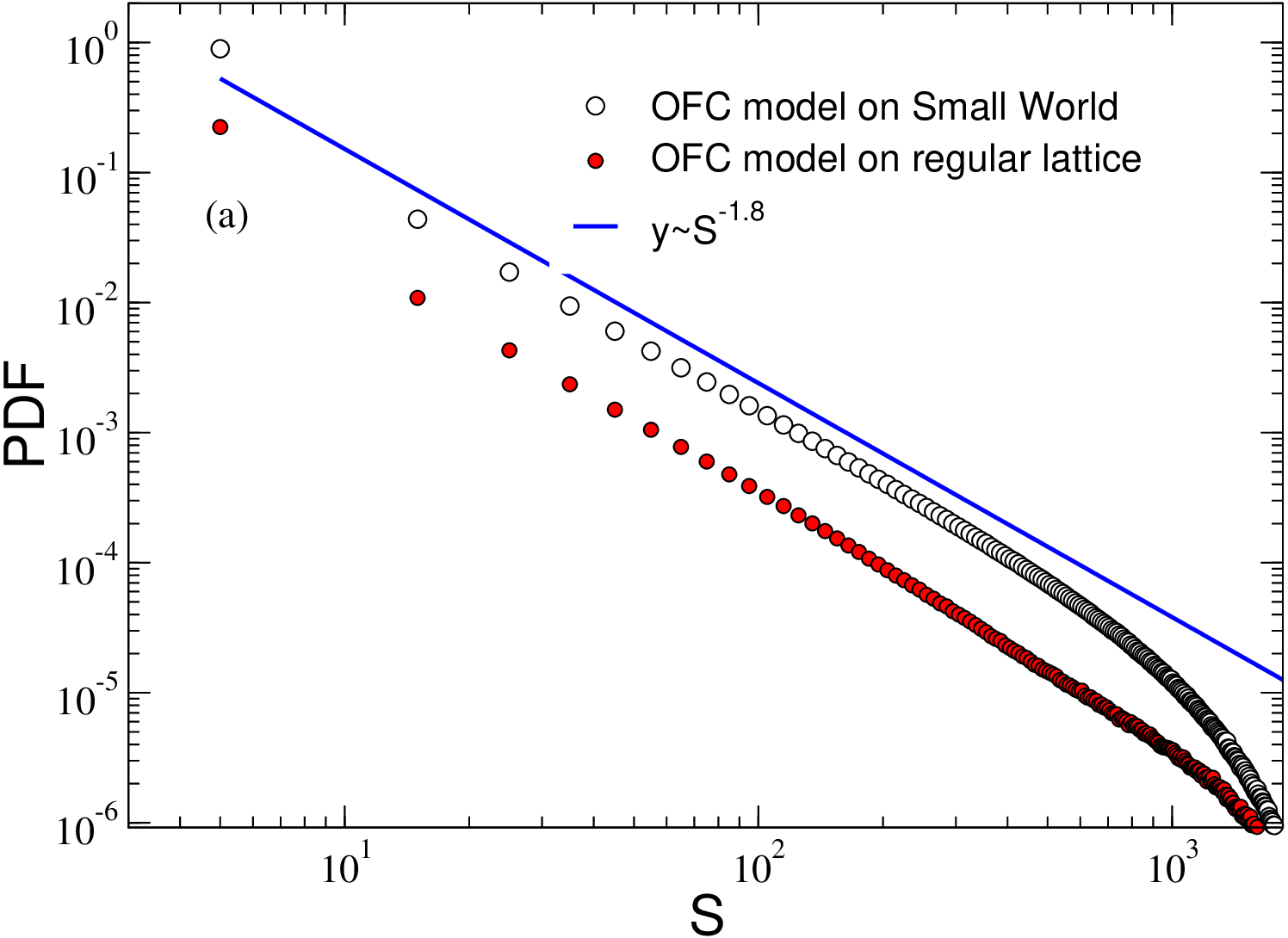}
\includegraphics[width=8cm]{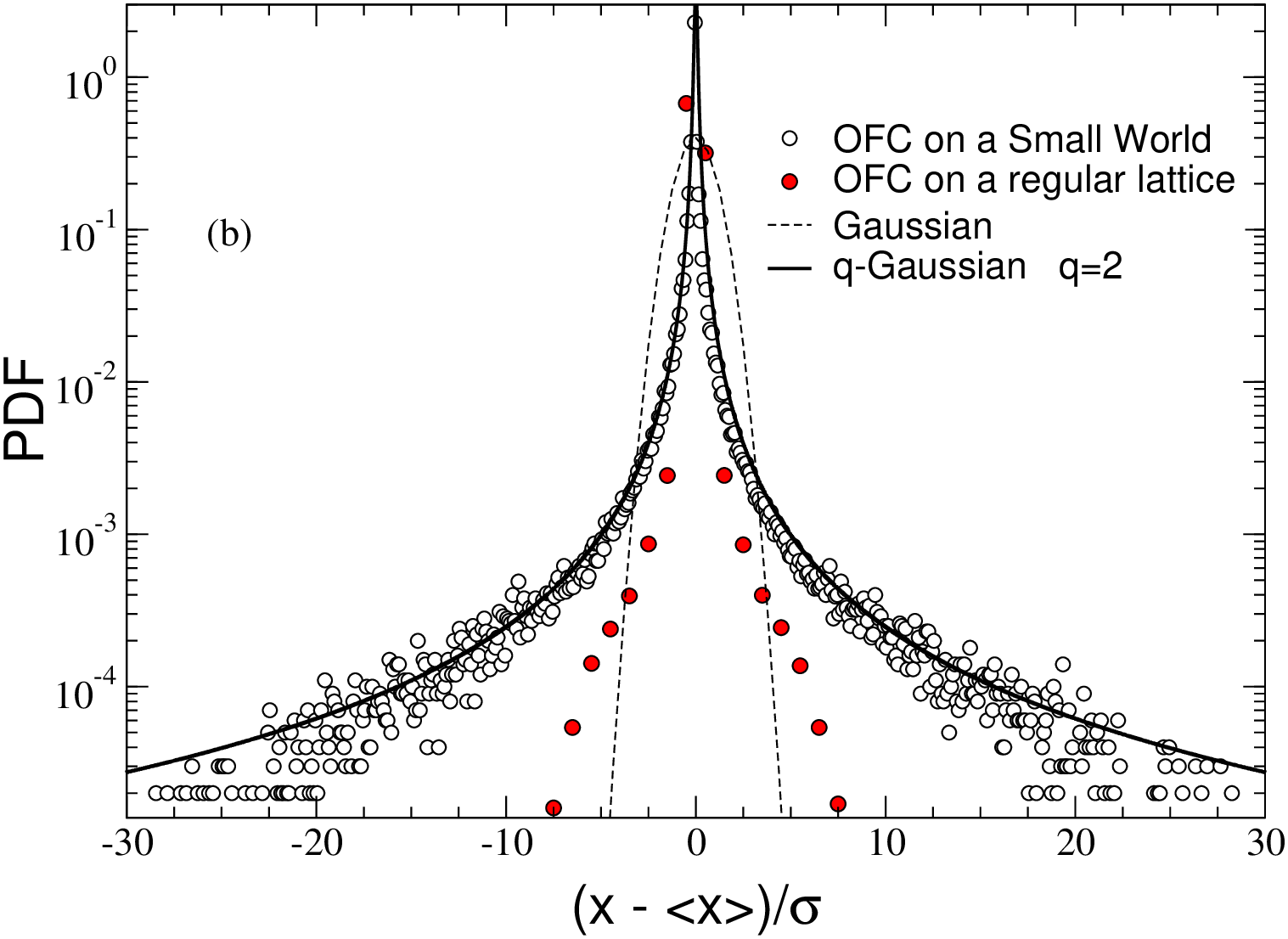}
\caption{\label{expM} (Color online) (a) Power-law distribution of
the avalanche size  $S$ for the OFC model ($\alpha=0.21$) on a
small world topology (open circles) and on a regular lattice 64x64
(full circles). Data were shifted for clarity. A fitting curve
with slope $\tau=1.8$ is also reported as full line. (b) PDF of
the avalanche size differences (returns) $x(t)=S(t+1)-S(t)$ for
the OFC model on a small world topology (critical state, open
circles) and on a regular lattice (non critical state, full
circles). Returns are normalized to the standard deviation
$\sigma$.  The first curve has been fitted with a q-Gaussian (full
line) with an exponent $q\sim2.0\pm0.1$. A standard Gaussian
(dashed line) is also reported for comparison. All the  curves
were normalized so to have unitary area. See text for further
details.}
\end{center}
\end{figure}
%
%
In Fig.1 (a) we plot the distribution of the avalanche size
time-series $S(t)$ for the OFC model on a small world topology (open circles)
and on a regular lattice (full circles). In our case the time
 $t$ is  a progressive discrete index
labelling successive events and is analogous to the   "natural
time" successfully used in \cite{varotsos}.
  We have considered up to $10^9$  avalanches to have  a good statistics.
In both cases the  data follow  a power-law decay $y\sim
S^{-\tau}$ with a slope $\tau=1.8\pm0.1$ even if criticality is
present only for the small world topology \cite{Filippo_creta}. In
the last years SOC models  have been  intensively studied
considering  time intervals between avalanches in the  critical
regime \cite{corral1}. Here we follow a different approach which
reveals interesting information on the eventual criticality  of
the model under examination. Inspired  by recent studies  on
turbulence and intermittent data \cite{frisch,stella,beck}, we
focus our attention  on the \emph{''returns''}
$x(t)=S(t+\Delta)-S(t)$, i.e. on the differences between avalanche
sizes calculated at time $t+\Delta$ and at time $t$, $\Delta$
being a  discrete time interval.

The resulting signal is extremely intermittent at
criticality, since successive events can have very different
sizes.
On the other hand, if  the system is not in a critical
state this intermittency character is very reduced. In Fig.1 (b)
we plot as open circles the Probability Density Function (PDF) of
the returns $x(t)$ (with $\Delta=1$) obtained for the critical  OFC
model on small world topology. The returns are normalized in order
to have zero mean and unitary variance. The curves  reported have
also unitary area. A behaviour very different from a Gaussian
shape (plotted as dashed curve) is observed. Data are  very peaked
with fat tails. On the other hand, for the model on regular lattice, even if
 power laws are found,  the model  is
not critical  since no finite-size scaling is observed
\cite{Filippo_creta}. In this case  no fat tails exist, although a sensible
departure from Gaussian behaviour is still present (see full
circles). These findings suggest a new powerful way for characterizing
the presence of criticality.
They are  also reinforced by similar results on other
 SOC  models not reported here for lack of space.
Another remarkable  feature is that such a behaviour does not
depend on the interval $\Delta$ considered for the  avalanche size
difference.  Also reshuffling the data, i.e. changing in a random way the
time order of the avalanches,  no change in the
PDFs is observed.   The data reported  in Fig.1 (b) for the critical OFC model
on a small world can be well fitted by  a \textit{q-Gaussian }
curve  $~~f(x)= A[1-(1-q){x^2}/B]^{1/(1-q)}~~$ typical of Tsallis
statistics\cite{tsallis}. This function generalizes the standard
Gaussian curve, depending on the parameters $A,B$ and on the
exponent $q$.  For  $q=1$ the normal distribution is obtained
again, so $q \ne 1$ indicates a departure from Gaussian
statistics. The q-Gaussian curve, reported as full line,
reproduces very well the model behaviour in the critical regime,
yielding in our case  a  value of $q=2.0\pm0.1$.
In order to compare these theoretical results with real data sets,
we repeated the previous analysis for the world wide seismic
catalog available on line \cite{wc}. We considered 689000
earthquakes in  the period 2001-2006. As a further term of
comparison, we selected a more complete seismic data set, i.e. the
Northern California catalog  for the period 1966-2006 \cite{nc}.
The latter is a very extensive seismic data set on one of the most
active and studied faults on the Earth, i.e. the San Andreas
Fault. In this case the total number of earthquakes is almost
400000.
As pointed out by several authors \cite{sornette} the energy, and
not the magnitude, is the quantity  which should be considered
equivalent to the avalanche size in the OFC model. In this paper
we consider the quantity $S=exp(M)$, $M$ being the magnitude of a
real earthquake. This quantity is simply related to the energy
dissipated in an earthquake, being the latter an increasing
exponential function of the magnitude.
\end{multicols}
\begin{figure} [ht]
\begin{center}
\includegraphics[width=8cm]{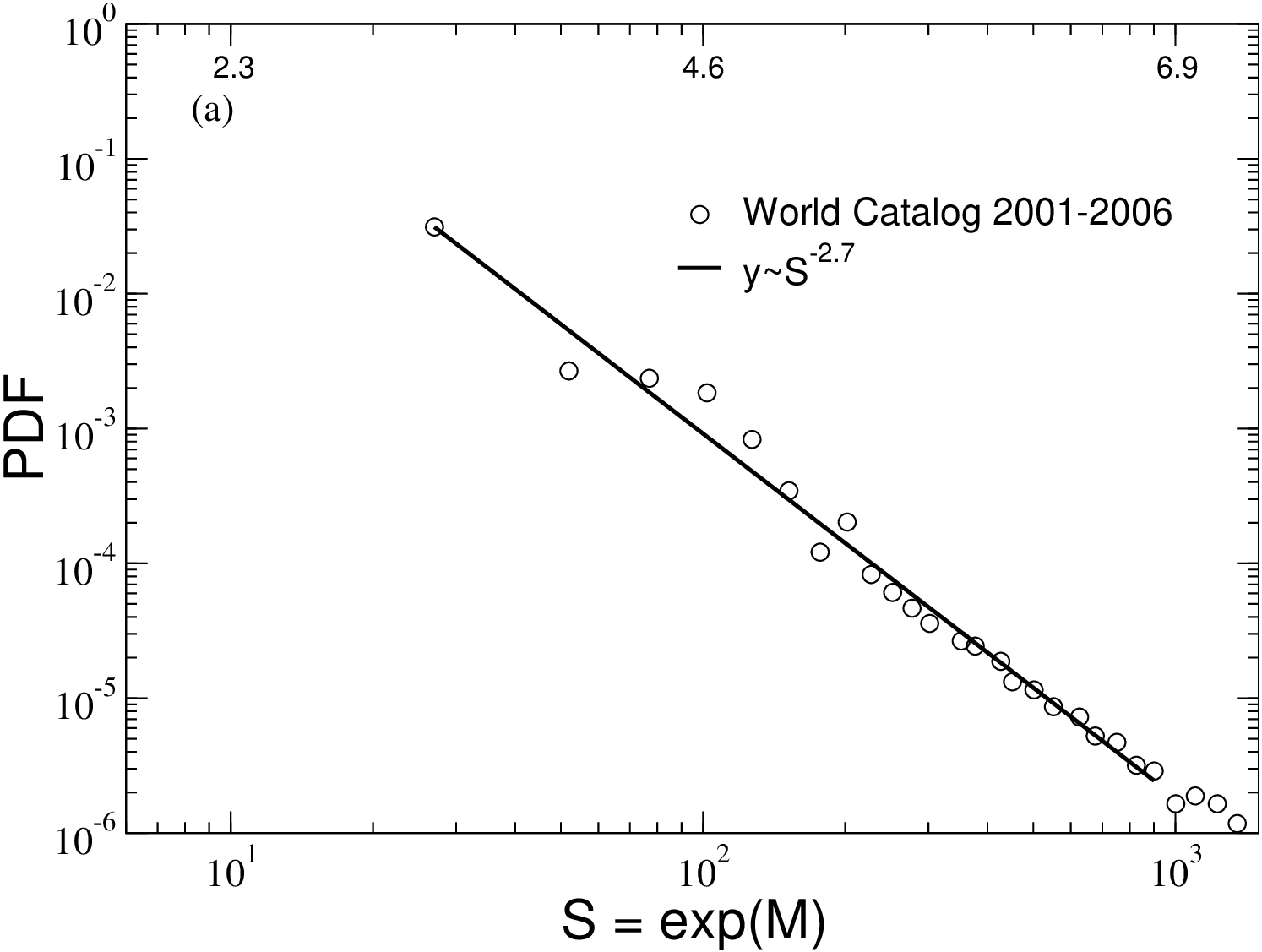}
\includegraphics[width=8cm]{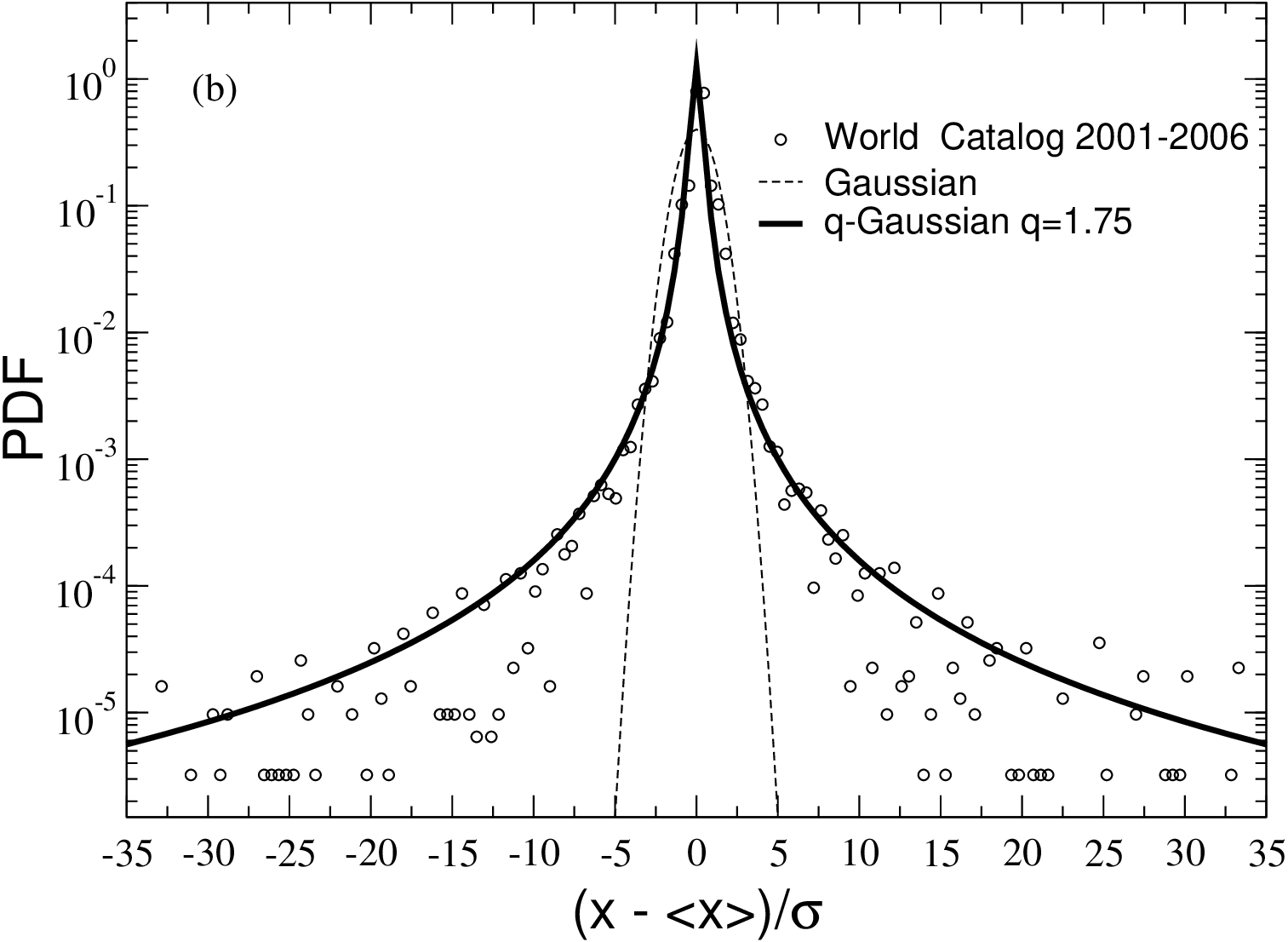}
\end{center}
\begin{center}
\includegraphics[width=8cm]{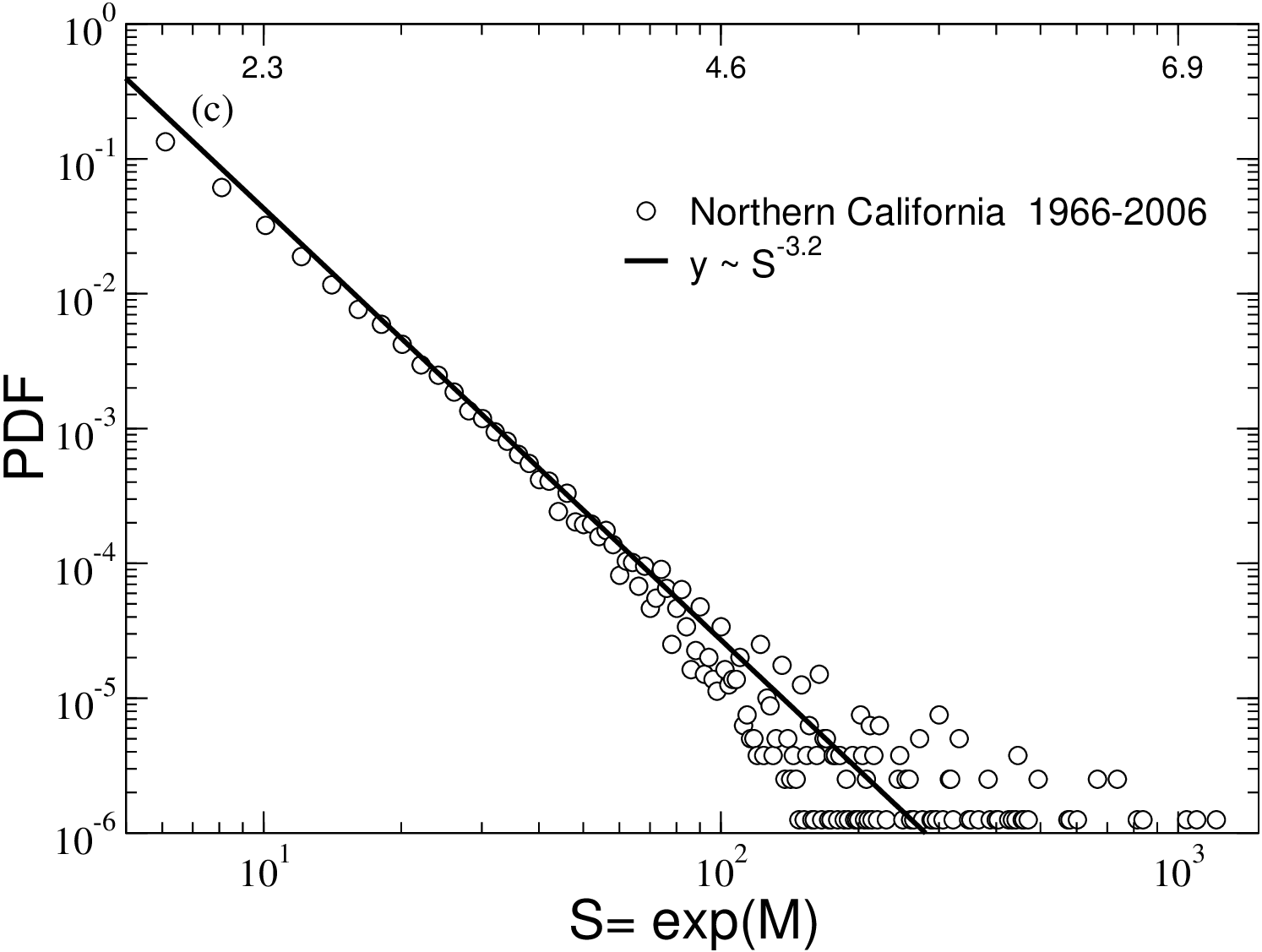}
\includegraphics[width=8cm]{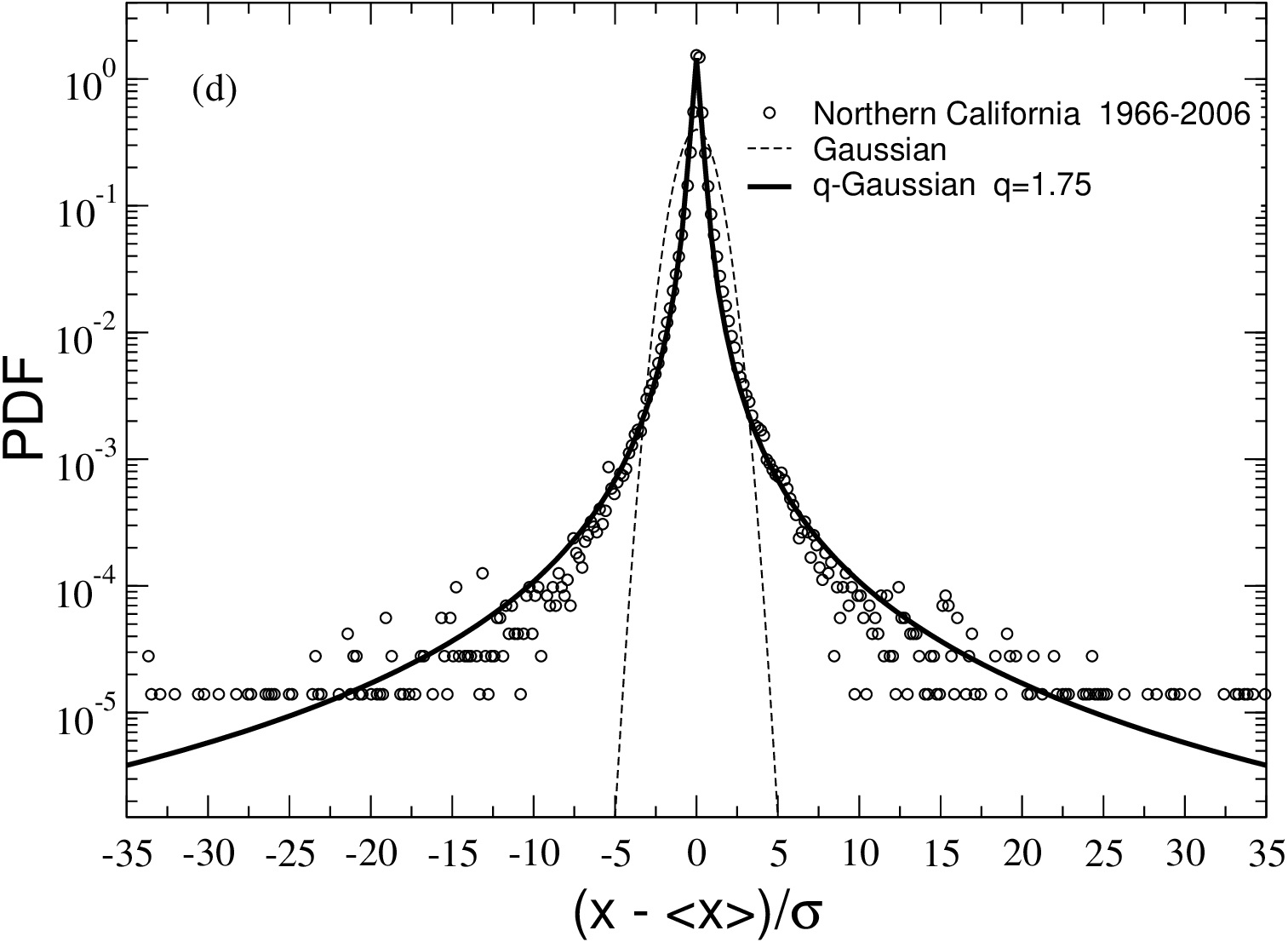}
\end{center}
\caption{\label{pdf-x} Power-law distribution of $S=exp(M) $, $M$
being the magnitude for the
World Catalog (a) and for the  Northern California catalog (c).
The correspondent power-law  fits are also reported. We show
the corresponding values of the magnitude M in the upper part of the figures.
PDFs of the energy differences $x(t)=S(t+1)-S(t)$ are
shown in (b) for the World Wide Seismic Catalog and in (d) for the
Northern California Catalog. In both the figures the data have
been fitted with a q-Gaussian (full line) with $q\sim1.75\pm0.15$.
A standard Gaussian is plotted as dotted line in all the figures
for comparison. See text for further details.}
\end{figure}
%
%
\begin{multicols}{2}
In Fig.2 (a) and (c)  we plot  the  PDFs of $S$ for the World
Catalog and the Northern California catalog. Power-law decay with
exponents $\tau=2.7\pm0.2$  and  $\tau=3.2\pm0.2$ reproduce the
PDFs  for the two cases respectively. Then we consider the PDFs of
the corresponding  returns $x(t)=S(t+\Delta)-S(t)$ (with
$\Delta=1$) and we plot them in Fig.2 (b) and (d).
Also for  real data \textit{t } is a progressive discrete index labelling
 successive events.
As  for the
critical OFC model previously discussed, fat tails and
non-Gaussian probability density functions are observed. In both
cases the experimental points can be fitted by a q-Gaussian curve,
obtaining an exponent $q=1.75\pm0.15$, a value which is
compatible, within the errors, to that one found for the OFC
model.
The  world  catalog  presents  large  fluctuations
in the tails due to the significant incompleteness given from the
lack of small magnitudes events at the global scale.
Also for the real earthquakes data, by changing the
interval $\Delta$ of the energy returns $x$, or by reshuffling the
time-series $S(t)$, no change in the  PDFs is observed.
In general both for the OFC and for  the real earthquakes catalogs the
avalanche  sizes (energies) $S$ occur with a power-law probability
$p(S)\sim S^{-\tau}$. In the hypothesis of no correlation between
the size of two events, the probability of obtaining the
difference $x=S'(t+\Delta)-S(t)$  (whatever $\Delta$) is given
by
\ba
P(x) &=& \textit{K} \int_{{0}}  ^\infty dS
\int_{0}^\infty dS' ~ (S S')^{-\tau} ~ \delta (S'  - S - x) = \nonumber \\
     &=& \textit{K} \int_{{\epsilon}} ^\infty  dS ~[S (S + |x|)]^{-\tau}
\ea
where \textit{K} is a normalization factor. The absolute value
$|x|$ takes into account the exchange of $S'$ with $S$. The
integral is divergent for $S=0$, so we  consider a small positive
value $\epsilon$ as an inferior limit of integration. Then one
gets
\ba P(x) = \textit{K} ~\frac{\epsilon^{-(2\tau-1)} }{ 2\tau-1}
~{_2F_1}(\tau,2\tau-1;2\tau; -\frac{|x|}{\epsilon})~~, \ea $_2F_1$
being the  confluent hypergeometric  function. The probability
density function (3) is plotted  for various values of $\tau$ in
Fig.3 (a). All these curves can be very well reproduced  by means
of q-Gaussians, whose values of $q$ do not depend on $\epsilon$
(since we have verified that the latter changes only the
normalization factor). The relation between $q$ and $\tau$ is
shown in Fig.3 (b), where the points are well fitted by a
stretched exponential curve. Notice that increasing $\tau$, i.e.
when  the power-law  tends  to an exponential, $q$ tends  to 1 as
expected. The value we get for the avalanche size power-law of the
OFC model with a small world topology is $\tau=1.8$, which
corresponds, according to Fig.3, to a value of $q\sim2.1$ in
agreement with the curve shown in Fig.1(b) within the errors. A
similar  correspondence can be found  for the  returns  of real
earthquakes data. In particular for  $\tau=2.7\pm0.2$  and
$\tau=3.2\pm0.2$, see Fig.2 (a)-(c),  one gets values of $q$
compatible, inside the errors, with the value $q\sim1.75$ found in
Fig.2 (b)-(d), see the values inside the box of Fig.3(b).
\begin{figure} [ht]
\begin{center}
\includegraphics[width=8cm]{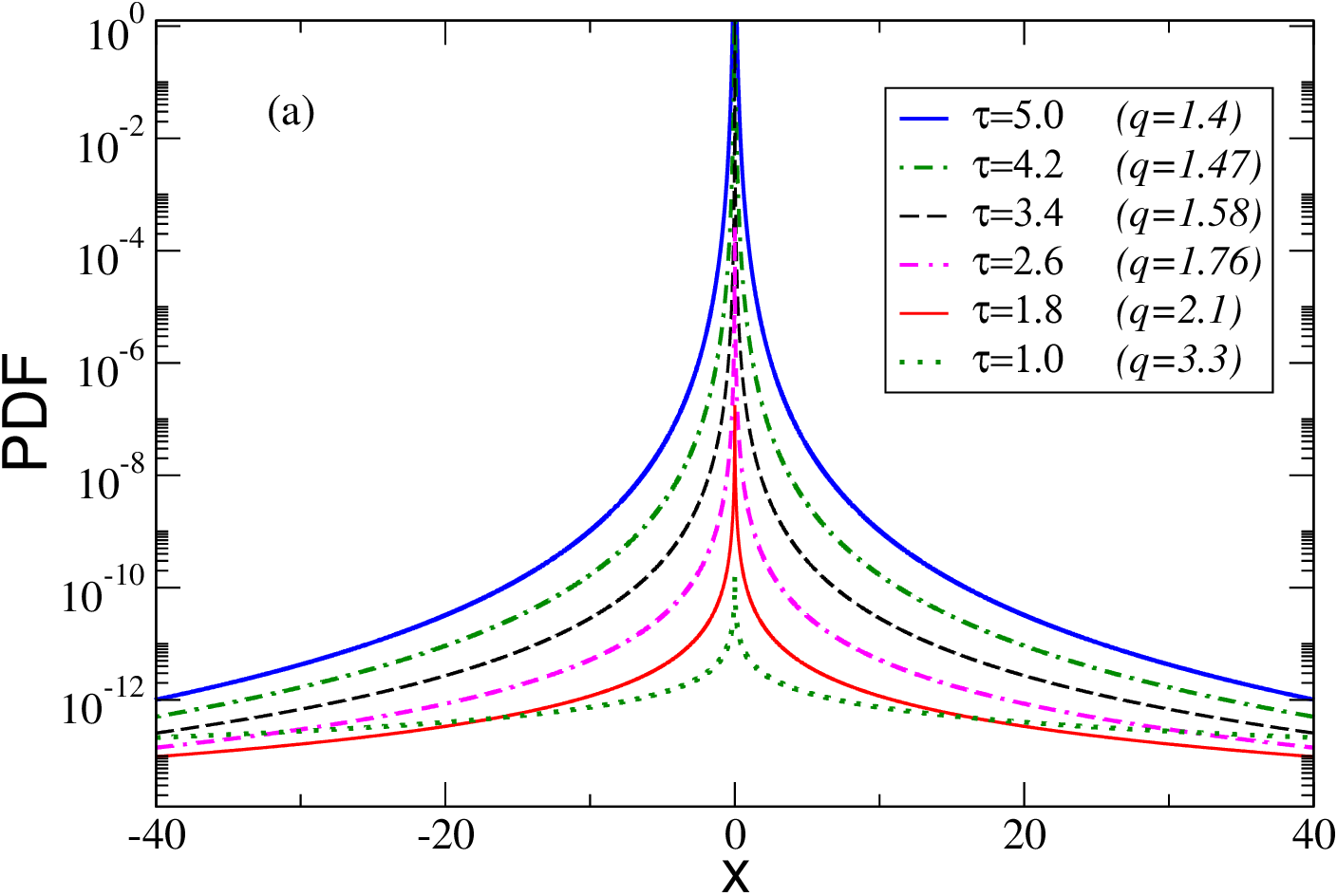}
\includegraphics[width=8cm]{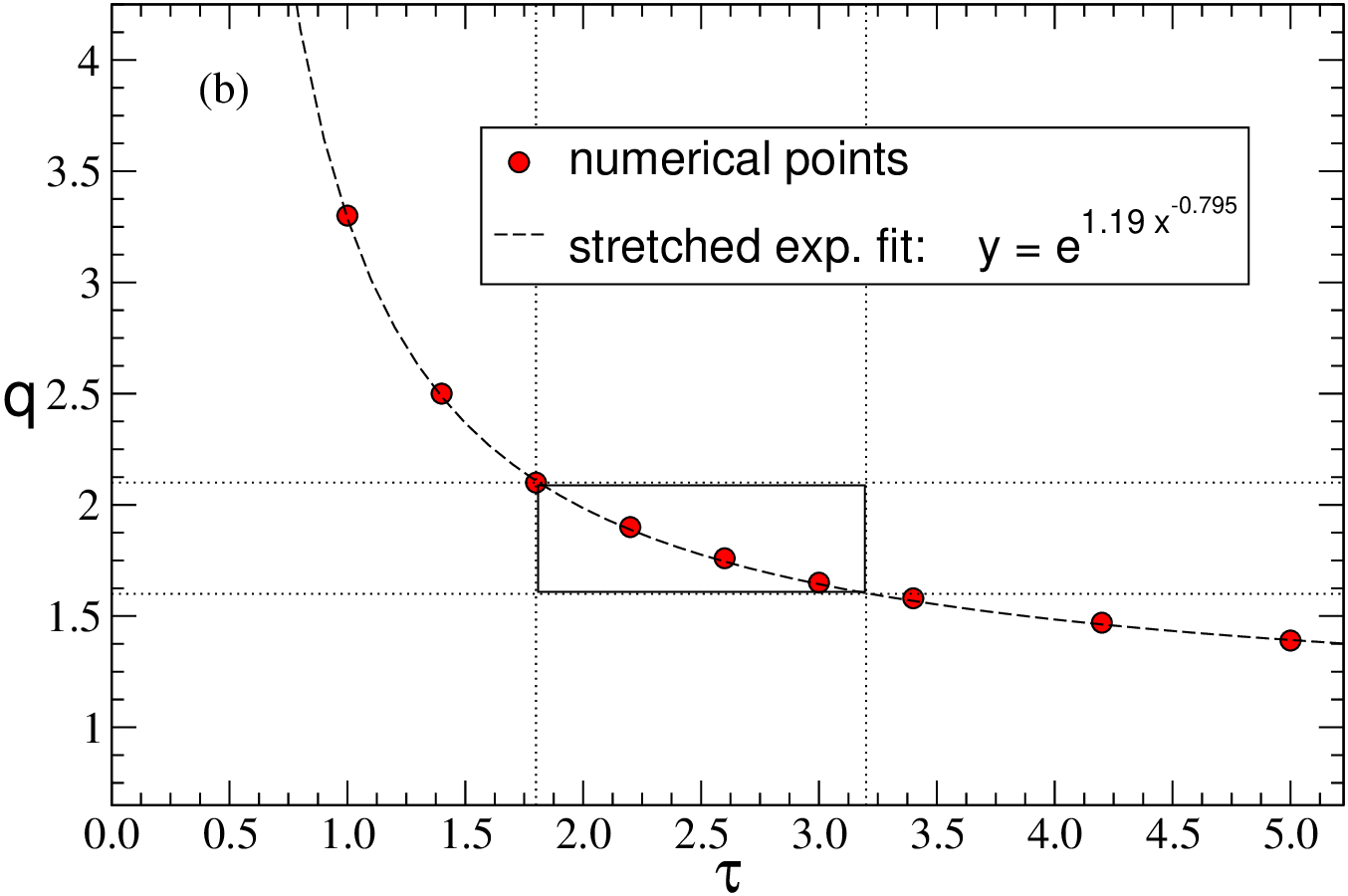}
\caption{\label{pdf-x} (Color online) (a) We  show the curves
obtained by calculating the integral of Eq.(3) considering
different values of $\tau$. (b) We plot (full circles) the
relationship between the values  of $ q$ and $\tau$, where $q$ is
obtained by fitting the curves in (a) with a q-Gaussian. The
resulting points can be very well reproduced by a stretched
exponential, $y=e^{1.19 x^{-0.795}}$, drawn as dashed line. Inside
the plotted box, one can find the values of  $\tau$ and $q$
 found  for the OFC model and the real data sets. }
\end{center}
\end{figure}
This result explains the q-Gaussian fat-tails in terms of
differences between uncorrelated (in time) avalanches
(earthquakes) sizes. On the other hand,  we have checked that when
avalanches are generated  by a deterministic chaotic dynamics, one
finds a dependence on the interval considered for the size returns
and the resulting PDFs cannot  be explained with the function (3).

In conclusion we have presented a  new analysis  which is able to
discriminate in a quantitative way real SOC  dynamics.  The
results here presented for the OFC model and   earthquakes data,
on one hand give further support to the hypothesis that seismicity
can be explained within a dissipative self-organized criticality
scenario when long-range interactions are considered. On the other
hand, although temporal and spatial correlations among avalanches
(earthquakes) do surely exist and a certain degree of statistical
predictability is likely possible, they indicate that it is not
possible to predict the magnitude of seismic events.

\textit{Acknowledgements:} We thanks S. Abe and  P.A. Varotsos for
useful discussions and comments. We acknowledge  financial support
from the PRIN05-MIUR project \textit{Dynamics  and thermodynamics
of systems  with long-range interactions. }

\end{multicols}


\begin{thebibliography}{99}
\bibitem{footnote}
We have checked that the results here discussed  are   valid
 for  $0.15\le \alpha \le 0.21$.
\end{thebibliography}
\end {document}